\documentclass[aps,showpacs,psfig,twocolumn,floatfix]{revtex4}
\usepackage{graphicx}
\usepackage{epsfig}
\begin{document}
\def\be{\begin{equation}}
\def\ee{\end{equation}}
\def\ba{\begin{eqnarray}}
\def\ea{\end{eqnarray}}

\title{
Interaction matrix element fluctuations in ballistic quantum dots:
random wave model}

\author{L. Kaplan$^1$ and Y. Alhassid$^2$}

\affiliation{$^1$Department of Physics, Tulane University, New Orleans,
Louisiana 70118, USA\\
$^{2}$Center for Theoretical Physics, Sloane Physics Laboratory, Yale
University, New Haven, CT 06520, USA}

\begin{abstract}
We study matrix element fluctuations of the two-body screened Coulomb
interaction and of the one-body surface charge potential in ballistic quantum
dots. For chaotic dots, we use a normalized random wave model to obtain analytic
expansions for matrix element variances and covariances in the limit of large
$kL$ (where $k$ is the Fermi wave number and $L$ the linear size of the dot). These leading-order analytical results are compared with exact numerical results. Both two-body and one-body matrix elements are shown to follow strongly non-Gaussian distributions, despite the Gaussian random nature of the single-electron wave functions.
\end{abstract}

\maketitle

\section{Introduction}

 There has been much interest in the  properties of quantum dots
whose single-particle dynamics are chaotic~\cite{alhassid00}.  The generic
fluctuation properties of the single-particle spectrum and wave functions in
such dots are usually described by random matrix theory (RMT)~\cite{guhr98}.
In open dots that are strongly coupled to leads, electrons can often be treated
as non-interacting quasi-particles, and the mesoscopic fluctuations of the
conductance have been explained using RMT.

 However, in almost-isolated dots, electron-electron interactions are important
and must be taken into account.  The simplest model of such dots is the constant interaction (CI) model, in which the interaction is taken to be the classical charging energy.  Charging energy leads to Coulomb blockade peaks in the conductance
versus gate voltage.  Each peak occurs as the gate voltage is tuned to
compensate for the Coulomb repulsion and an additional electron tunnels into
the dot.  For a fixed number of electrons, the CI model is essentially a
single-particle model, and RMT can be used to derive the statistical properties
of the conductance peak heights~\cite{jalabert92}.  While the CI plus RMT model
has explained (at least qualitatively)~\cite{jalabert92,alhassid96,alhassid98}
several observed features of the peak height fluctuations~\cite{folk96,chang96,folk01}, there have been significant discrepancies with experimental data, in particular regarding the peak spacing
statistics~\cite{sivan96,simmel97,patel98a,luscher01}.  Such discrepancies indicate the importance of interactions beyond charging energy.

 A more systematic way of treating electron-electron interactions in chaotic
ballistic dots is to expand the interaction in a small parameter, the inverse
of the Thouless conductance $g_T \sim k L$, where $k$ is the Fermi wave number
and $L$ is the linear size of the dot.  The Thouless conductance measures the
number of single-particle levels within an energy window determined by the time
it takes the electron to cross the dot, and $g_T$ increases as the square root of the
number of electrons.  It can be shown that, in the limit of large Thouless
conductance, only a few interaction terms survive, constituting the interacting
part of the universal Hamiltonian~\cite{kurland00,aleiner02}.  These universal
interaction terms include, in addition to charging energy, a constant exchange
interaction.  The inclusion of an exchange interaction has explained the
statistics of peak heights at low and moderate temperatures as well as the
suppression of the peak spacing fluctuations~\cite{alhassid03,usaj03}.
However, at low temperatures, the peak spacing distribution remains bimodal
even when the exchange interaction is included, while none of the experimental
distributions are bimodal~\cite{sivan96,patel98a,simmel97,luscher01}.

 For finite Thouless conductance, residual interactions beyond the universal
Hamiltonian must be taken into account.  The randomness of the single-particle
wave functions induces randomness in the two-body screened Coulomb interaction
matrix elements~\cite{blanter97}. The possible induced two-body ensembles have
been classified according to their underlying space-time symmetries and features
of the two-body interaction~\cite{alhassid05}. In a Hartree-Fock-Koopmans~\cite{koopmans34} approach (assuming the Hartree-Fock single-particle wave functions do not change as electrons are added to the dot), the peak spacing can be expressed directly in terms of certain diagonal interaction matrix elements~\cite{alhassid02}. Sufficiently large fluctuations of these interaction matrix elements can explain the absence of bimodality in the peak spacing distribution~\cite{alhassid02,usaj02}.  The variance of these fluctuations is determined by the spatial correlations of the single-particle wave functions.  In a diffusive dot these correlations have been derived to leading order in $1/g_T$, and the variance of the matrix elements of the screened Coulomb interaction was shown
to behave as $\Delta^2/g_T^2$~\cite{blanter97,mirlin00}, where $\Delta$ is the
mean level spacing of the single-particle spectrum.  However, dots studied in the
experiments are usually ballistic. Wave function correlations in ballistic
quantum dots are much less understood.  Berry's conjecture~\cite{berry77}
regarding the Gaussian nature of wave function fluctuations in chaotic systems
provides the leading order behavior of the correlations at short distances, but
finite-size contributions at distances that are comparable to the size of the
dot can have important effects on the matrix element fluctuations.

 An additional contribution to the peak spacing fluctuations originates in
surface charge effects~\cite{blanter97}.  In a finite-size system, screening
leads to the accumulation of charge on the surface of the dot.  The confining
one-body potential is then modified upon the addition of an electron to the
dot.  In a diffusive dot, the variance of one-body matrix elements behaves as
$\Delta^2/g_T$.

Another interesting phenomenon, in which interaction matrix element fluctuations play an important role is spectral scrambling as electrons are added to a chaotic or diffusive dot~\cite{scrambling,alhassid07}. Both the two-body interaction and one-body surface charge effects are responsible for scrambling.

 Here we investigate fluctuations of the two-body interaction matrix
elements and of the surface charge potential matrix elements in ballistic dots.
We use a normalized version of Berry's random wave model to derive analytically leading-order contributions for an arbitrary dot geometry, and compare with exact numerical simulations.

 The outline of this paper is as follows.  In Sec.~\ref{random-wave} we review
the random wave model for chaotic billiards.  The spatial correlator of wave function intensity obtained from this model is geometry-independent but not
consistent with the normalization requirement of the wave
functions~\cite{gornyi02}.  We compute normalization corrections to the
correlator in Sec.~\ref{seccorrel}. The variances of diagonal,
double-diagonal, and off-diagonal two-body interaction matrix elements are calculated in Sec.~\ref{two-body} and are all found to be strongly affected by the
normalization correction. As a result, ratios of these variances are shown to remain far from their asymptotic $kL \to\infty$ values for the range $30 \le kL \le 70$ that is relevant to experiments. In Sec.~\ref{covar} we study the covariance
of interaction matrix elements, relevant for understanding spectral scrambling
when several electrons are added to the dot~\cite{scrambling}, and supplement the full random wave analysis with a schematic random matrix model.  One-body matrix element variances are treated in Sec.~\ref{one-body}.  In all of these
studies we compare numerical results of the random wave model with
leading-order analytical estimates. In Sec.~\ref{secdistr} we study the interaction matrix element distributions and show that, in the experimentally accessible range of $kL$, they deviate significantly from the Gaussian limit implied by the central limit theorem.  Finally, in Sec.~\ref{summary} we address additional implications of the present work, including the necessity to supplement the normalized random wave model
with dynamical effects for quantitative comparison with experiment.

\section{Random Wave Model}\label{random-wave}

 In a chaotic system without symmetries, a typical classical trajectory will
uniformly explore an entire energy hypersurface in phase space.  A
well-established and extensively tested conjecture by Berry~\cite{berry77}
holds that in the quantization of such a system, a typical wave function will
 spread uniformly over an energy hypersurface, up to the inevitable
Gaussian random amplitude fluctuations that arise whenever a random vector is
expanded in a generic basis.  In a semiclassical picture, the Gaussian
fluctuations are the result of exponentially many wave fronts visiting any
given region of phase space with quasi-random phases.

 For a two-dimensional billiard system, the random wave model implies that a
typical chaotic wave function may be written locally as a random superposition
of plane waves at fixed energy $\hbar^2k^2/2m$:
\begin{equation}
\label{plexp}
\psi(\vec r)=\psi(r,\theta)= \int_0^{2\pi} d\phi\,
  A(\phi)e^{ikr \cos(\theta-\phi)} \,,
\end{equation}
where $A(\phi)$ is distributed as a $\delta$-correlated Gaussian random
variable:
$\overline{A(\phi)}=0$ and $\overline{ A^\ast(\phi) A(\phi')}
= {1 \over 2 \pi V} \delta(\phi-\phi')$.
The normalization is fixed by $\overline{|\psi(\vec r)|^2}={1/V}$, where $V$ is
the area of the billiard.  Equivalently, the wave function may be expanded in
circular waves with good angular momentum $\mu$:
\begin{equation}
\label{besselexp}
\psi(\vec r)=\sum_{\mu=-\infty}^\infty B_\mu \phi_\mu(\vec r) \,,
\end{equation}
where
\begin{equation}\label{basis}
\phi_\mu(\vec r) \equiv J_\mu(kr) e^{i \mu \theta} \;.
\end{equation}
The discrete variables $B_\mu$ ($\mu = 0, \pm 1, \pm 2, \ldots$) are taken to be
uncorrelated Gaussian random variables with
\begin{equation}\label{B-mu}
\overline {B_\mu} =0 \;\;\;\;\;\; \overline{B^\ast_{\mu} B_{\mu'}} = {1 \over
V} \delta_{\mu\mu'}.
\end{equation}
 The normalization integral of a wave function (\ref{besselexp}) is given by
\begin{equation}
{1 \over V} \int_V d\vec r \, |\psi(\vec r)|^2 = \sum_{\mu \mu'} B_\mu^\ast
A_{\mu \mu'} B_{\mu'} \,,
\end{equation}
where $A_{\mu \mu'}$ are the basis state overlaps
\begin{equation}
A_{\mu\mu'} = {1 \over V} \int_V d \vec r \, \phi^\ast_\mu(\vec r)
\phi_{\mu'}(\vec r)\;.
\label{adef}
\end{equation}
 Defining ${\bf B}$ to be a column vector with components $B_\mu$, the
normalization integral of the wave function $\psi$ is simply the norm of ${\bf
B}$ with the matrix $A$ playing the role of a metric
\begin{equation}\label{norm}
{1\over V}\int_V d\vec r \,|\psi(\vec r)|^2 = {\bf B}^\dagger A {\bf B} \;.
\end{equation}

 The random coefficients $B_\mu$ may be chosen to either satisfy or not satisfy the
real wave function condition $B_{-\mu}=B^\ast_\mu$, corresponding to the
absence or presence of an external magnetic field; these two situations are
conventionally denoted by $\beta=1$, $2$ respectively.

 Using the addition theorem for Bessel functions we have
\begin{equation}
\sum_{\mu=-\infty}^\infty \phi^\ast_\mu(\vec r) \phi_{\mu}(\vec r') =
J_0(k|\vec r-\vec r'|) \,.
\label{besselident}
\end{equation}
 It follows from the completeness property (\ref{besselident}) of the Bessel
basis that
\begin{equation}\label{trace}
{\rm tr}\; A =\sum_\mu A_{\mu\mu}= 1 \;.
\end{equation}

 We note that formally the model requires an infinite set of basis states; in
practice the effective number $N_{\rm eff}$ of basis states that have
appreciable magnitude inside the area $V$ scales as $N_{\rm eff} \sim k L \sim
g_T$, where $L\equiv \sqrt V$ is a typical linear size of the billiard and
$g_T$ is the ballistic Thouless conductance.  A more accurate estimate for
$N_{\rm eff}$ can be obtained by considering a disk of radius $R=L/\sqrt{\pi}$.
In a disk, the wave functions $\phi_\mu$ are orthogonal because of rotational
symmetry and $A_{\mu \mu'} \propto \delta_{\mu \mu'}$.  The effective dimension
$N_{\rm eff}$ is then obtained as the ``participation number" of the exact
$A_{\mu\mu}$:
\begin{equation}
N_{\rm eff}= \left[\sum_\mu |A_{\mu\mu}|^2 \right]^{-1} \approx 1.85 kR
\approx 1.04 kL \;.
\label{neff}
\end{equation}
 The approximate completeness of a basis of $N_{\rm eff} \sim g_T$ plane waves
at a fixed energy $\hbar^2k^2/2m$ was confirmed in studies of a billiard
system~\cite{murthy05}.

\section{Intensity Correlator}\label{seccorrel}

 The random wave model of Eqs.~(\ref{besselexp}) and (\ref{B-mu}) together with
Eq.~(\ref{besselident}) leads to the amplitude correlator
\begin{equation}
\overline{\psi^\ast(\vec r) \psi(\vec r')}= {1 \over V} J_0(k|\vec r- \vec r'|)
\label{ampcorr}
\end{equation}
and intensity correlator
\begin{eqnarray}
& &\overline{|\psi(\vec r)|^2 |\psi(\vec r')|^2} -\overline{|\psi(\vec r)|^2}
\;\overline{|\psi(\vec r')|^2} \nonumber \\
& & \approx C(\vec r,\vec r') = {1 \over V^2} {2 \over \beta} J_0^2(k|\vec r-
\vec r'|) \;. \label{corr}
\end{eqnarray}
 Eq.~(\ref{corr}) is obtained from Eq.~(\ref{ampcorr}) by contracting
$\psi(\vec r)$ with $\psi^\ast(\vec r')$ and $\psi^\ast(\vec r)$ with
$\psi(\vec r')$ and noting that there are two equivalent ways of performing
the contraction when $\psi=\psi^\ast$ (i.e., for $\beta=1$).

 Similar correlators can be derived starting from the plane-wave expansion
(\ref{plexp}).  The intensity correlator (\ref{corr}) is valid to leading order
in $|\vec r - \vec r'| /L$, but becomes problematic when applied to to all
$\vec r$, $\vec r'$ in the finite area $V$.  Indeed wave function
normalization requires the correlator to vanish on average~\cite{mirlin00},
\begin{equation}\label{normalization}
\int_V d \vec r \; \left[\overline{|\psi(\vec r)|^2 |\psi(\vec r')|^2}
-\overline{|\psi(\vec r)|^2} \;\overline{|\psi(\vec r')|^2} \right]   =  0\,,
\end{equation}
and similarly $\int_V d \vec r' \, [\ldots ]=0$.  However, the random wave
intensity correlator $C(\vec r, \vec r')$ in (\ref{corr}) is non-negative
everywhere and does not satisfy the condition (\ref{normalization}).  The
reason for this failure is that in the random wave model normalization is
satisfied only on average, i.e., $\overline{\int_V d \vec r \, |\psi(\vec r)|^2
}= {\rm tr}\, A = 1$, but not for each individual wave function.

 This deficiency can be corrected by introducing the {\em normalized} random
wave model, in which each ``random'' wave function (\ref{besselexp}) is
normalized in area $V$, i.e.,
\begin{equation}
\psi^{\rm norm}(\vec r) = \sum_\mu B^{\rm norm}_\mu J_\mu(k r) e^{i\mu\theta}
\end{equation}
with $B^{\rm norm}_\mu  =
{ B_\mu / \sqrt{V\; {\bf B}^\dagger A {\bf B}}}$.
The normalized random wave model is easy to implement
numerically by normalizing each random wave.  Analytically, the intensity correlator of this model can be written as
\begin{eqnarray}
C^{\rm norm}(\vec r,\vec r')
\!=\! \overline{
\left (\!|\psi^{\rm norm}(\vec r)|^2 \!-\!{1\over  V} \!\!\right ) \!\!\!
\left (\!|\psi^{\rm norm}(\vec r')|^2 \!-\!{1\over V} \!\!\right )} \,.
\label{cnorm1}
\end{eqnarray}
This is similar to the situation in random matrix theory (RMT), where the naive
guess $\overline{|a_i|^2|a_j|^2}-\overline{|a_i|^2}\;\overline{|a_j|^2} = {1
\over N^2}\delta_{ij}$ for $\beta=2$ ($a_i$ and $a_j$ are two components of an
RMT eigenvector of length $N$ with normalization $\sum_i |a_i|^2=1$) must be
replaced by the exact expression ${1 \over N(N+1)}\left [\delta_{ij}-{1 \over
N}\right]$ to obtain correct normalization for finite $N$.  In our case,
however, the normalized intensity correlator (\ref{cnorm1}) depends on both
positions $\vec r$ and $\vec r'$ as well as the system geometry.

We define at a spatial point ${\vec r}$ an excess wave-function intensity by
$u(\vec r)=|\psi(\vec r)|^2-{1 \over V}$
and similarly an excess normalized wave-function intensity
$u^{\rm norm}(\vec r)= |\psi^{\rm norm}(\vec r)|^2-{1 \over V}$.
We then have
\begin{eqnarray}
\label{cnorm}
&C^{\rm norm}(\vec r,\vec r')=\overline{u^{\rm norm}(\vec r)
u^{\rm norm}(\vec r') }  &\nonumber \\[6pt]
&= \overline{ \left ({{1 \over V}+ u(\vec r) \over 1+\int_V d\vec r_a \, u(\vec
r_a)} -{1 \over V} \right ) \left ({{1 \over V}+ u(\vec r') \over 1+\int_V
d\vec r_b \, u(\vec r_b)} -{1 \over V} \right ) } \,. &
\end{eqnarray}

 Eq.~(\ref{cnorm}) is exact but unwieldy.  In the semiclassical limit of large
$kL$, a given superposition $\psi$ of random waves will be almost normalized,
i.e.,  $\int_V d{\vec r} \, |\psi(\vec r)|^2-1=\int_V  d{\vec r} \, u(\vec r)
=O((kL)^{-1/2})$.  Thus, to leading order in $1/kL$, we may expand
Eq.~(\ref{cnorm}) to obtain
\begin{eqnarray}
\overline{
\left ( u(\vec r) - {1 \over V} \!\int_V \! d{\vec r_a} \, u(\vec r_a)
\right)\!\! \left ( u(\vec r') - {1 \over V} \!\int_V  \! d{\vec r_b} \, u(\vec
r_b) \right) \!+\cdots} \nonumber \,, \\ \;
\end{eqnarray}
where we have omitted all terms involving
three-point and higher-order correlations of the excess intensity $u$.
Eq.~(\ref{cnorm}) can be simplified to obtain
\begin{equation}
\label{cnormexp}
C^{\rm norm}(\vec r,\vec r')= \tilde C (\vec r,\vec r')+ O\left ({ 1 \over
(kL)^{3/2}}\right) \,,
\end{equation}
where
\begin{eqnarray}
\tilde C(\vec r,\vec r') &=& C(\vec r,\vec r') - {1 \over V}
\int_V d\vec r_a \, C(\vec r,\vec r_a) \nonumber \\&-& {1 \over V}
\int_V d\vec r_a \, C(\vec r_a,\vec r') \nonumber \\&+& {1 \over V^2} \int_V
\int_V d\vec r_a d\vec r_b \,C(\vec r_a,\vec r_b) \,.
\label{corrnorm}
\end{eqnarray}

 The leading-order normalized correlator $\tilde C(\vec r,\vec r')$ was derived
in Ref.~\cite{gornyi02} by adding a weak smooth disorder and using the
non-linear supersymmetric sigma model. More recently, the same leading correction
has been obtained~\cite{urbina07} starting from a density matrix and
the principle of maximum entropy~\cite{goldstein}.
Here we we have shown that this form
can be obtained quite generally by applying a perturbative approach to the
unnormalized correlator $C(\vec r,\vec r')$, assuming only that correlations
are weak in the limit of large system size $kL$.

We note, however, that
$\tilde C(\vec r,\vec r')$ is merely a leading order approximation in $1/kL$ to
the true normalized random wave correlator $C^{\rm norm}(\vec r,\vec r')$, as
it involves only terms depending on the unnormalized two-point correlator $C$.
The complete expression (\ref{cnorm}) for the normalized two-point correlator
$C^{\rm norm}(\vec r,\vec r')$ involves all unnormalized $n$-point correlators
$\overline{u(\vec r_1) \cdots u(\vec r_n)}$. In particular, the unnormalized
three-point correlator gives rise to the $O((kL)^{-3/2})$ correction in
Eq.~(\ref{cnormexp}).  In principle, such higher-order corrections to
Eq.~(\ref{cnormexp}) may be computed systematically, but in practice their
effect on matrix element variances is small, as we will confirm below.  The $n$-point correlators $\overline{u(\vec r_1) \cdots u(\vec r_n)}$ for $n\ge 3$ will be
important when we discuss the deviation of the matrix element distribution from a Gaussian (see Sec.~\ref{secdistr}).
We also note that our definition of $C^{\rm norm}(\vec r,\vec r')$ and the approach of Refs.~\cite{urbina07,goldstein} may produce different corrections to $\tilde C(\vec r,\vec r')$ at higher orders in a $1/kL$ expansion. Although the problem of constructing the ``best" random wave ensemble that exactly satisfies normalization constraints is an interesting one in its own right, the effect of these higher-order ambiguities on matrix element statistics is negligible compared to dynamical contributions in real chaotic systems.

In contrast to the unnormalized correlator $C(\vec r, \vec r')$, the normalized correlator $\tilde C(\vec r, \vec r')$ satisfies
\begin{equation}
\int_V d\vec r' \, \tilde C(\vec r,\vec r')=0 \,,
\label{tildecint}
\end{equation}
implying the normalization of each individual wave function $\int_V d\vec r \, |\psi_\alpha(\vec r)|^2 =1$.
To show this, we define ${\cal
N}_\alpha = \int_V d\vec r \, |\psi_\alpha(\vec r)|^2$. We note that   $\overline{\cal N}=1$ by construction,
and calculate the variance in wave function normalization:
\begin{eqnarray}
\overline{{\cal N}^2}- \overline{{\cal N}}^2  &=&
\int_V  \int_V d\vec r \, d\vec r' \,
\overline{|\psi(\vec r)|^2 |\psi(\vec r')|^2}-1 \nonumber
\\ &=& \int_V  \int_V d\vec r \, d\vec r' \left ( {1 \over V^2} + \tilde C(\vec
r,\vec r') \right ) -1 \nonumber
\\ &=& \int_V  \int_V d\vec r \, d\vec r' \,\tilde C(\vec r,\vec r') \,.
\end{eqnarray}
Eq.~(\ref{tildecint}) implies that the variance of
 ${\cal N}$ vanishes, and since the average wave function
is normalized to unity, every wave function in the ensemble must be normalized to
unity.

 A more schematic random matrix approximation is obtained if the exact metric
$A_{\mu \mu'}$ is assumed to be diagonal and the diagonal components are
replaced by
\begin{eqnarray}\label{a-rmt}
A_{\mu \mu} = \left\{ \begin{array}{ll} 1/N_{\rm eff}& {\rm for} \;\; |\mu| <
N_{\rm eff}/2 \\ 0 & {\rm for} \;\; |\mu| \geq N_{\rm eff}/2
 \end{array} \right.
\end{eqnarray}
 The distribution (\ref{a-rmt}) satisfies $\sum_\mu A_{\mu\mu}=1$ and $\sum_\mu
|A_{\mu\mu}|^2 =1/N_{\rm eff}$ and therefore reproduces correctly the first two
moments (\ref{trace}) and (\ref{neff}) of the exact $A_{\mu\mu}$ distribution.
In the approximation (\ref{a-rmt}), the normalization of the wave function reads
${\bf B}^\dagger {\bf B} = N_{\rm eff}/V$ [see Eq.~(\ref{norm})].  The
normalized $N_{\rm eff}$-component vector ${\bf \tilde B} = (N_{\rm
eff}/V)^{-1/2} \,{\bf B}$ can then be viewed as an eigenvector of a Gaussian
random matrix of the corresponding symmetry class $\beta$.

\begin{figure}[ht] \begin{center} \leavevmode \parbox{0.5\textwidth}
{
\psfig{file=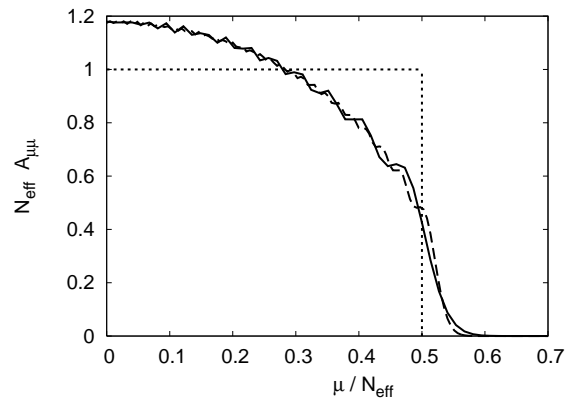,width=0.3\textwidth,angle=270}
}
\end{center}
\protect\caption{
The exact Bessel function self-overlaps $A_{\mu\mu}$ defined by
Eq.~(\ref{adef}) are computed for a disk of radius $R$, and scaled by the
effective Hilbert space dimension $N_{\rm eff}=1.85 kR$.  The solid curve
indicates $kR=40$, while the dashed curve corresponds to $kR=100$.  The dotted
rectangle represents the schematic random matrix approximation, where all
$A_{\mu\mu}$ are taken to be constant for $-N_{\rm eff}/2 <\mu <N_{\rm eff}/2$
and $0$ otherwise.  Quantities plotted in this and all subsequent figures
are dimensionless.
}
\label{figamatr}
\end{figure}

 In a disk of radius $R$, asymptotic expressions may be used to verify that the
exact self-overlaps of the Bessel functions approach a constant $A_{\mu\mu}
\approx 2/\pi k R$ for $\mu \ll N_{\rm eff} \sim kR$ and fall off exponentially
as $A_{\mu\mu} \approx {1 \over 2\pi\mu^2} \left({ek R\over 2
\mu}\right)^{2\mu}$ for $\mu \gg N_{\rm eff} \sim kR$.  In Fig.~\ref{figamatr},
we show the exact random-wave metric for a disk with $kR=40$ and $kR=100$, as
well as the schematic random matrix approximation (\ref{a-rmt}).

 In the following sections we use the random wave model to estimate the
variances of various matrix elements of the screened Coulomb interaction.  Normalized random waves can be generated numerically to calculate the spatial wave function correlator and the corresponding variances, which is equivalent to using the exact normalized intensity correlator $C^{\rm norm}$.  Analytical expressions can be obtained as follows.  (i) The asymptotic behavior of the unnormalized
random wave correlator $C(\vec r,\vec r')$ in Eq.~(\ref{corr}) will be
sufficient to produce analytically the leading geometry-independent
$O\left({\ln kL / (kL)^2}\right)$ behavior of two-body interaction matrix
element variances. (ii) A direct integration of the normalized correlator
$\tilde C(\vec r, \vec r')$ in Eq.~(\ref{corrnorm}) produces geometry-dependent
$O\left({1 /(kL)^2}\right)$ terms for the above matrix elements.  This
correlator can also be used to calculate to leading order in $kL$ the size of
one-body surface charge matrix element fluctuations.

\section{Two-body interaction matrix elements}\label{two-body}

 Here we model the screened two-body Coulomb interaction in 2D quantum dots as
a contact interaction $v(\vec r, \vec r')=  \Delta\,V \delta(\vec r-\vec r')$,
where the single-particle mean level spacing $\Delta$ serves to set the
energy scale~\cite{altshuler97,ullmo}.

\subsection{Fluctuation of diagonal matrix elements $v_{\alpha\beta}$}
\label{randvab}

 We first discuss the variance of a diagonal two-body matrix element
$v_{\alpha\beta} \equiv v_{\alpha\beta;\alpha\beta}$.  In the contact
interaction model
\begin{equation}
v_{\alpha\beta}=\Delta V \int_V d\vec r \, |\psi_\alpha(\vec r)|^2
|\psi_\beta(\vec r)|^2 \;.
\label{contactint}
\end{equation}
To leading order in $g_T \sim kL$, the dominant
contribution to the variance arises from correlations between the intensities
of a single wave function at different points~\cite{scrambling}:
\begin{eqnarray}
\label{vabint}
\overline{\delta v_{\alpha\beta}^2}
= \Delta^2 V^2 \int_V \int_V d\vec r \, d\vec r' \,
\tilde C^2(\vec r,\vec r') +\cdots
\end{eqnarray}
Since this leading-order effect comes from correlations within a single wave
function, the only restriction on the energy difference $E_\alpha-E_\beta$
between the two random wave functions is $|E_\alpha-E_\beta| \ll E_\alpha$, or
equivalently $|k_\alpha-k_\beta| \ll k_\alpha$, allowing the variance to be a
function of a single parameter $kL$.  The subleading $O(1/(kL)^3)$ terms
omitted in Eq.~(\ref{vabint}) involve the two-eigenstate intensity correlator
\begin{equation}
\tilde C_2(\vec r,\vec r')=\overline{|\psi_\alpha(\vec r)|^2 |\psi_\beta(\vec
r')|^2}
- \overline{|\psi_\alpha(\vec r)|^2} \; \overline{|\psi_\beta(\vec r')|^2} \,.
\label{twoeig}
\end{equation}
The correlator (\ref{twoeig}) produces the leading effect for the covariance and will be discussed in detail in Section~\ref{covar}.

 The leading-order contribution to Eq.~(\ref{vabint}) is obtained by using the
unnormalized correlator $C(\vec r, \vec r')$ instead of $\tilde C(\vec r, \vec
r')$.  Changing integration variables in Eq.~(\ref{vabint}) to $\vec R=(\vec r
+\vec r')/2$ and $\vec \rho = \vec r -\vec r'$, inserting the unnormalized
correlator (\ref{corr}), which depends only on $k\rho$, and substituting the
asymptotic form $J_0^2(k \rho) = {2 \over \pi k\rho} \cos^2(k\rho-\pi/4)
+O\left(1/(k\rho)^{2}\right)$, we obtain
\begin{eqnarray}
\overline{\delta v_{\alpha\beta}^2} &\approx & {\Delta^2 \over V} \left({2
\over \beta} \right)^2  \int_{1/k}^L d\rho\, (2\pi\rho) \,
{ 4 \over \pi^2k^2 \rho^2}\cos^4(k\rho-\pi/4) \nonumber \\
& = &
\Delta^2 {3 \over \pi}
\left({2 \over \beta} \right)^2 {\ln kL \over (kL)^2} \,,
\label{leadingvabrand}
\end{eqnarray}
where in the last line we have used $\overline{\cos^4{k \rho}}=3/8$.

 In Eq.~(\ref{leadingvabrand}) we have not properly included the short distance
contribution from $\rho \sim 1/k$ and the shape-dependent long-distance
contribution from $\rho \sim L$. Both of these contributions to Eq.~(\ref{vabint})
scale as $1/k^2$; in the
first case because $\rho \sim 1/k$ defines an $O(1/k^2)$ volume in $\vec
\rho$-space, and in the second case because $J_0^4(k \rho) \sim 1/(kL)^2$ for
$\rho \sim L$.  To obtain the correct result at this subleading order we need
to use the normalized correlator of Eq.~(\ref{corrnorm}).  We then obtain
\begin{eqnarray}
\label{vabint2}
&\!\!\!\!\!\!\!\!\!\!\!\!\!
\overline{\delta v_{\alpha\beta}^2} & \!= \! \Delta^2 V^2 \!\!
\int_V \int_V \! d\vec r \, d\vec r' \,
\tilde C^2(\vec r,\vec r') +O\left({\Delta^2 \over (kL)^3}\right) \\
&& \!=\! \Delta^2 {3 \over \pi} \left({2 \over \beta} \right)^2 {\ln kL
+b_g\over (kL)^2} + O\left({\Delta^2 \over (kL)^3}\right) \,.
\label{bcoeff}
\end{eqnarray}
The leading $\ln kL/(kL)^2$ term, discussed in Ref.~\cite{ullmo},
depends only on the symmetry class,  while the shape-dependent coefficient $b_g$ can be easily evaluated by numerical integration of Eq.~(\ref{vabint2}).

\begin{figure}[ht] \begin{center} \leavevmode \parbox{0.5\textwidth}
{
\psfig{file=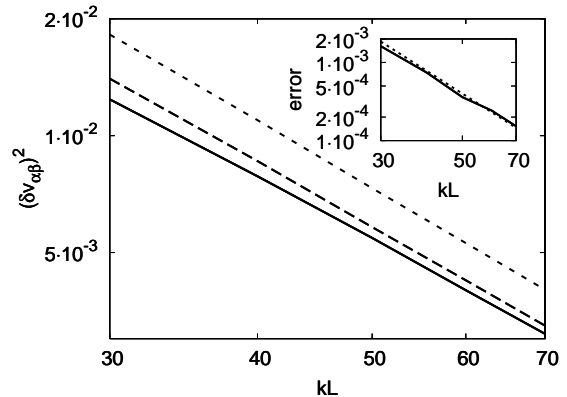,width=0.3\textwidth,angle=270}
}
\end{center}
\protect\caption{
The variance of the two-body matrix element $v_{\alpha\beta}$ versus $kL$ in
the real ($\beta=1$) random wave model: (a) the solid curve is the result of
exact numerical simulations; (b) the long-dashed line is the result of integrating
the square of the normalized correlator $\tilde C^2(\vec r,\vec r')$, as in
Eq.~(\ref{vabint2}); (c) the short-dashed line is the result of using the
unnormalized correlator $C^2(\vec r,\vec r')$.  In the inset, the solid line is
the difference between the full result (a) and the approximation (b); the
dotted line indicates that omitted terms scale as ${1 \over (kL)^3}$.  Here and
in all following figures, the level spacing $\Delta$, which sets the overall
energy scale, is set to unity.}
\label{figrandvab}
\end{figure}

 In Fig.~\ref{figrandvab} we show the results for a disk geometry in the range
$30 \le kL \le 70$, corresponding roughly to the parameter range relevant for
experiments ($\sim 150 - 800$ electrons in the dot).  The numerical simulation of $\overline{\delta v_{\alpha\beta}^2}$ (solid line) is compared with the leading result of Eq.~(\ref{vabint2}) (long-dashed line).  For the disk, we find $b_g=-0.1$.  The difference between the full numerical calculation and the integral of the
normalized correlator is plotted in the inset, and is seen to
be in good agreement with the $1/(kL)^3$ scaling of
Eq.~(\ref{vabint2}).  If the unnormalized correlator (\ref{corr})
is used in (\ref{vabint2}) we find (for the disk geometry) $b^{\rm
unnorm}_g=0.9$.  The corresponding variance is shown as a short-dashed line in
Fig.~\ref{figrandvab}.

\begin{figure}[ht] \begin{center} \leavevmode \parbox{0.5\textwidth}
{
\psfig{file=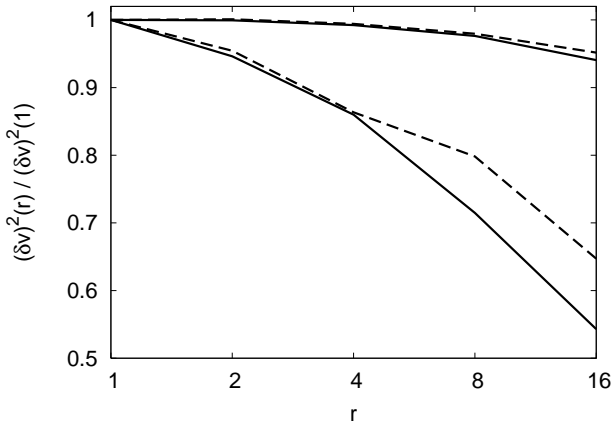,width=0.33\textwidth,angle=270}
}
\end{center}
\protect\caption{
The change in the $v_{\alpha\beta}$ variance (upper two curves) and
the variance of the one-body matrix element $v_{\alpha}$ (lower two curves, see Sec.~\ref{one-body})
for normalized real random waves are plotted
as functions of the aspect ratio $r=a/b$ of the elliptical geometry. Here
$kL=k\sqrt{\pi a b}$ is kept fixed at $30$ (solid curves) and $70$ (dashed curves).
}
\label{figrandvabshape}
\end{figure}

 The dependence of the $v_{\alpha\beta}$ fluctuations on the dot's geometry is
demonstrated in the upper two curves Fig.~\ref{figrandvabshape}.  Here we plot the result of
Eq.~(\ref{vabint2}) for elliptical shapes as a function of the aspect ratio $r$
between the major and minor axes, while keeping the area and the wave vector
$k$ fixed.  We find that as the aspect ratio changes by a (physically
unrealistic) factor of $16$, the shape-dependent parameter $b_g$ only changes
from $-0.1$ to $-0.3$, resulting in a change of only $\sim 5-6\%$ in the
$v_{\alpha\beta}$ variance for the experimental range of $kL$.  Similar results
are found if elliptical shapes are replaced by rectangles or other geometries.
Thus, for all practical purposes, shape effects on the $v_{\alpha\beta}$
variance can be ignored (at least within the normalized random wave model).

At this subleading order, we must in principle also take into account the
energy difference $E_\alpha-E_\beta$, if this energy difference is larger than
the ballistic Thouless energy $g_T \Delta$ (but still small compared with the
average $(E_\alpha+E_\beta)/2$).  Then $1/L \ll\delta k= k_\alpha-k_\beta \ll
k= (k_\alpha+k_\beta)/2$, and the $\cos^4(k\rho-\pi/4)$ factor in
Eq.~(\ref{leadingvabrand}) must be replaced by
$\cos^2(k_\alpha\rho-\pi/4)\cos^2(k_\beta\rho-\pi/4)$, which has the average
value $3/8$ only for $1/k \ll \rho \ll 1/\delta k$ and is reduced to an average
of $1/4$ for $1/\delta k \ll \rho  \ll L$.  After performing the integration
over $\rho $ in Eq.~(\ref{leadingvabrand}), we find that $\ln kL$ in the final
result must be replaced with $\ln kL - (1/3)\ln \delta k L$.  Assuming $\delta
k L$ remains large but fixed while $kL \to \infty$, this corresponds merely to
a modification of the geometry-dependent coefficient in Eq.~(\ref{bcoeff}):
$b_g \to b_g - (1/3) \ln \delta k L$ for $\delta k L \gg 1$.  In practice, for
reasonable energy windows, e.g., $\delta k L \sim 5$, the
consequent reduction in the variance is at most $10\%$
and may be safely ignored compared to the much larger dynamical effects
present in real chaotic systems.

\subsection{Fluctuation of $v_{\alpha\alpha}$ and
$v_{\alpha\beta\gamma\delta}$}
\label{secrandvaa}

 The preceding section studied fluctuations in the intensity overlap
$v_{\alpha\beta}$ between two random wave functions $\psi_\alpha$ and
$\psi_\beta$.  Similar techniques may be applied to the self-overlap of a
single wave function $v_{\alpha\alpha}=\Delta \,V \int_V d\vec r \,
|\psi_\alpha(\vec r)|^4$, also known as the inverse participation ratio in
coordinate space, and to the ``off-diagonal" four-wave-function overlap
$v_{\alpha\beta\gamma\delta}=\Delta \,V \int_V d\vec r \, \psi^\ast_\alpha(\vec
r)\psi_\beta(\vec r)\psi^\ast_\gamma(\vec r)\psi_\delta(\vec r)$.

 To leading order, $O\left({\ln kL / (kL)^2}\right)$, all the results arise
from integrating the unnormalized correlator (\ref{corr}) and differ only by
combinatorial factors.  For example, the $({2 \over \beta})^2$ factor in
Eq.~(\ref{leadingvabrand}) may be understood from the fact that there are four
distinct ways to contract pairs of same-wave-function amplitudes between
$\psi^\ast_\alpha(\vec r) \psi_\alpha(\vec r) \psi^\ast_\beta(\vec r)
\psi_\beta(\vec r)$ and $\psi_\alpha(\vec r') \psi^\ast_\alpha(\vec r')
\psi_\beta(\vec r') \psi^\ast_\beta(\vec r')$
if the wave functions are real ($\beta=1$), but
only one way to perform this contraction if the wave functions are complex
($\beta=2$). An analogous counting argument for the variance of $v_{\alpha\alpha}$
leads to
\begin{equation}
\label{vaaint2}
\overline{\delta v_{\alpha\alpha}^2}
= \Delta^2 {3 \over \pi} c_\beta {\ln kL +b'_g\over (kL)^2} +
O\left({\Delta^2 \over (kL)^3}\right)
\end{equation}
with $c_1=24$ and $c_2= 4$.  We note that such combinatorial factors can also
be derived from the invariance of the second moments of the matrix elements
under a change of the single-particle basis~\cite{alhassid05}.

 Finally, if $\alpha$, $\beta$, $\gamma$, and $\delta$ are all different, there
is only one way to perform the contraction in either the real or complex case,
leading to
\begin{eqnarray}\label{vabcdint2}
\overline{\delta v_{\alpha\beta\gamma\delta}^2}
& = & \Delta^2 V^2 \int_V \int_V
d\vec r \, d\vec r' \left[\overline {\psi^\ast(\vec
r)\psi(\vec r')}\right]^4 \nonumber \\  & = &\Delta^2 {3 \over \pi} {\ln kL
+b''_g\over (kL)^2} +
O\left({\Delta^2 \over (kL)^3}\right) \,.
\end{eqnarray}

 While the leading ${\ln kL / (kL)^2}$ behavior in
Eqs.~(\ref{vaaint2}) nd (\ref{vabcdint2}) is identical to that for
$\overline{\delta v_{\alpha\beta}^2}$ up to geometry-independent combinatorial
factors, the coefficients of the subleading geometry-dependent $1/(kL)^2$ terms
are not so simply related, as indicated by $b'_g, b''_g \ne b_g$ in the above
expressions.  That is because the normalization-related subtraction, which
enters at $O\left({1 / (kL)^2}\right)$, works differently for the three
matrix elements.  For example, the variance of $v_{\alpha\beta}$ involves the
product of two normalized intensity correlators $\tilde C^2(\vec r,\vec r')
\sim (J_0(k|\vec r-\vec r'|)^2 - \cdots)^2$.  On the other hand, the variance
of $v_{\alpha\beta\gamma\delta}$ involves the product of four amplitude
correlators (i.e., $\overline {\psi^\ast(\vec r)\psi(\vec r')}^4 \sim
J_0(k|\vec r-\vec r'|)^4$), which do not require subtraction.  As indicated
earlier, the absence of subtraction in the integral results in $b''_g=0.9$ for
a disk geometry, in contrast with $b_g=-0.1$.  For $v_{\alpha\alpha}$
fluctuations, we find $b'_g=-2.25$ for a disk~\cite{geometry}.
Thus, although $\overline{\delta v^2_{\alpha\alpha}}$ and $\overline{\delta
v^2_{\alpha\beta\gamma\delta}}$ are related to $\overline{\delta
v^2_{\alpha\beta}}$ by geometry-independent universal factors in the $kL \to
\infty$ limit, the convergence to this universal limit is logarithmically slow.
Specifically, in the presence of time-reversal symmetry ($\beta=1$),
\begin{eqnarray}
\overline{\delta v^2_{\alpha\alpha}} / \overline{\delta v^2_{\alpha\beta}} &=&6
+ {b'_g-b_g \over \ln kL} +\cdots \label{vaaratio} \\
\overline{\delta v^2_{\alpha\beta}} / \overline{\delta
v^2_{\alpha\beta\gamma\delta}} &=&4 + {b_g-b''_g \over \ln kL} +\cdots \,,
\label{vabgdratio}
\end{eqnarray}
to leading order in $1/\ln kL$.

\begin{figure}[ht] \begin{center} \leavevmode \parbox{0.5\textwidth}
{
\psfig{file=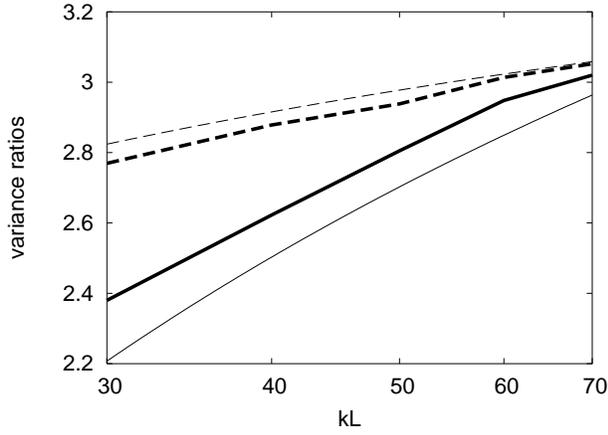,width=0.33\textwidth,angle=270}
}
\end{center}
\protect\caption{
Ratios of matrix element variances are plotted for real random waves inside a
disk as functions of $kL$: $\overline{\delta v^2_{\alpha\alpha}} /
\overline{\delta v^2_{\alpha\beta}}$ (thick solid line) and $\overline{\delta
v^2_{\alpha\beta}} / \overline{\delta v^2_{\alpha\beta\gamma\delta}}$ (thick
dashed line).  For comparison, the leading-order analytic results of
Eqs.~(\ref{vaaratio}) and (\ref{vabgdratio}) are shown as thin solid and dashed
lines, respectively.
}
\label{figrandratios}
\end{figure}

 Fig.~\ref{figrandratios} shows these ratios versus $kL$ in the
physical range of interest.  Note in particular the very slow convergence of
Eq.~(\ref{vaaratio}) to the asymptotic value of $6$ because of the large difference
between the coefficients of the subleading terms $b'_g - b_g = -2.15$.

\subsection{Matrix element covariance $\overline{\delta v_{\alpha \beta}
\delta v_{\alpha\gamma}}$}
\label{covar}

 Another quantity of interest is the covariance $\overline{\delta
v_{\alpha\beta} \delta v_{\alpha\gamma}}$, where $\psi_\alpha$, $\psi_\beta$,
and $\psi_\gamma$ are three distinct wave functions of a single dynamical
system.  Such matrix element covariances are important in quantitative
estimates of scrambling of the Hartree-Fock single-particle spectrum as
electrons are added to the dot~\cite{scrambling}.  As in Eq.~(\ref{vabint}),
the leading contribution for large $kL$ can be written as a double integral of
a product of two intensity correlators:
\begin{eqnarray}
\overline{\delta v_{\alpha\beta} \delta v_{\alpha\gamma}} \approx
 \int_V \int_V d\vec r \, d\vec r' \, \tilde C(\vec r,\vec r') \tilde
C_2(\vec r, \vec r') \;, \label{covarexpr}
\end{eqnarray}
where $\tilde C(\vec r, \vec r')$ is the intensity correlator for wave function $\psi_\alpha$ and $\tilde C_2(\vec r, \vec r')$ is the intensity correlator (\ref{twoeig}) between two distinct wave functions $\psi_\beta$ and $\psi_\gamma$.  In a diffusive dot, expression (\ref{covarexpr}) leads to a covariance $\propto
\Delta^2/g_T^3$, where $g_T$ is the diffusive Thouless conductance
\cite{scrambling}.

 We proceed to evaluate $\tilde C_2$ in the normalized random wave
approximation.  Clearly, if $\psi_\beta$, $\psi_\gamma$ are chosen to be
independent random wave functions, the correlator vanishes by construction.
Therefore, we need to impose the orthogonality condition
\begin{equation}
{1\over V} \int_V d\vec r \,
\psi_\beta^\ast(\vec r) \psi_\gamma(\vec r)= {\bf B}^\dagger_\beta A {\bf
B}_\gamma = 0 \,, \end{equation}
which may be done through the Gram-Schmidt procedure. We begin by generating two independent random vectors $\psi_\beta$ and $\psi_\gamma^0$ at the same energy $\hbar^2k^2/2m$: $\psi_\beta(\vec r) = \sum_{\mu=-\infty}^\infty B_{\beta\mu} J_\mu(kr) e^{i \mu \theta} \nonumber$ and  $\psi_\gamma^0(\vec r) = \sum_{\mu=-\infty}^\infty
B_{\gamma\mu} J_\mu(kr) e^{i \mu \theta}$,
project $\psi_\gamma^0$ onto the subspace orthogonal to $\psi_\beta$,
and normalize the result to unit length to find $\psi_\gamma$:
\begin{equation}
\psi_\gamma(\vec r)= { \psi_\gamma^0(\vec r) - \langle \psi_\beta
|\psi_\gamma^0\rangle\psi_\beta(\vec r) \over \sqrt{1-|\langle
\psi_\beta|\psi_\gamma^0\rangle|^2}} \,.
\end{equation}
Using the basis states $\phi_\mu$ in Eq.~(\ref{basis}), we find
\begin{eqnarray}
\tilde C_2(\vec r,\vec r') = {1 \over V^2}{2 \over \beta}
\left[J_0^2(k|\vec r-\vec r'|) \sum_{\mu\mu'}
|A_{\mu\mu'}|^2\right. \nonumber \\ - \! 2J_0(k|\vec r-\vec r'|)
\left. \sum_{\mu\mu'} A_{\mu\mu'} \phi_\mu(\vec r) \phi^\ast_{\mu'}(\vec r')
\right] \!+O\! \left({1 \over k^2}\right) \,. \label{correldiff}
\end{eqnarray}

 Eq.~(\ref{correldiff}) can be evaluated numerically for a given geometry in
order to obtain the covariance of diagonal matrix elements $v_{\alpha\beta}$
via Eq.~(\ref{covarexpr}).  However, we obtain insight into the qualitative
behavior of the correlator (\ref{correldiff}) by using the schematic random
matrix model (\ref{a-rmt}) in a disk geometry.  We find
\begin{equation}
\label{c2rmt}
\tilde C_2(\vec r,\vec r') \approx -{1 \over V^2}{2 \over \beta}{1 \over N_{\rm
eff}} J_0^2(k|\vec r-\vec r'|) = -{1 \over N_{\rm eff}}\tilde C(\vec r,\vec r')
\;,
\end{equation}
where $N_{\rm eff}$ is the effective dimension defined by (\ref{neff}).
Substituting the random matrix model relation (\ref{c2rmt}) in
(\ref{covarexpr}) and using Eq.~(\ref{vabint}), we have
\begin{equation}
\label{covarrmt}
\overline{\delta v_{\alpha\beta} \delta v_{\alpha\gamma}} \approx -{1 \over
N_{\rm eff}} \; \overline{\delta v_{\alpha\beta}^2} \,.
\end{equation}

 This simple relation between the variance and covariance of the interaction
matrix elements could also be obtained directly from an effective completeness
condition (valid for any $\alpha$)
\begin{equation}\label{completeness}
 \sum_{\beta=1}^{N_{\rm eff}} \delta v_{\alpha\beta}=0 \,,
\end{equation}
where $\delta v_{\alpha\beta}= v_{\alpha\beta} - \overline{v_{\alpha\beta}}$.
Eq.~(\ref{completeness}) follows from the completeness of the eigenstate basis
$\sum_\beta |\psi_\beta(\vec r)|^2 = c$ (where $c$ is a position-independent
constant).  From Eq.~(\ref{completeness}) we find
$\sum_{\beta \gamma} \delta v_{\alpha\beta} \delta v_{\alpha\gamma} =0$ or,
after taking an ensemble average
\begin{equation}\label{covar-var}
\sum_{\beta \neq \gamma} \overline{\delta v_{\alpha\beta} \delta
v_{\alpha\gamma} } = -\sum_\beta  \overline{(\delta v_{\alpha\beta})^2} \;.
\end{equation}
 If we assume the various covariances and variances to be independent of the
specific orbitals, we recover the RMT relation (\ref{covarrmt}).

 Substituting Eqs.~(\ref{neff}) and (\ref{bcoeff}) into Eq.~(\ref{covarrmt}),
we obtain an analytic result for the covariance in a circular dot in the RMT approximation:
\begin{equation}
\overline{\delta v_{\alpha\beta} \delta v_{\alpha\gamma}} \approx
-\Delta^2 {3 \over \pi} {\sqrt{\pi}\over 1.85}
\left({2 \over \beta} \right)^2 {\ln kL +b_g\over
(kL)^3} + O\left({\Delta^2 \over (kL)^4}\right) \,.
\label{rmtapprox}
\end{equation}

 For a non-circular dot, we can write
\begin{equation}
\phi_\mu(\vec r)\phi^\ast_{\mu'}(\vec r')= A^\ast_{\mu\mu'}J_0(k|\vec r-\vec
r'|) +f_{\mu\mu'}(\vec r,\vec r')\,,
\end{equation}
where from Eq.~(\ref{besselident}) $\sum_\mu f_{\mu\mu} =0 $ for arbitrary
$\vec r$, $\vec r'$, and from Eq.~(\ref{adef}) $\int_V d\vec r \;
f_{\mu\mu'}(\vec r,\vec r)=0$ for arbitrary $\mu$, $\mu'$.  Although we have no
formal justification for neglecting $f$, if we do so we obtain
\begin{equation}
\tilde C_2(\vec r,\vec r') \approx
-\left(\sum_{\mu\mu'} |A_{\mu\mu'}|^2 \right)\tilde C(\vec r,\vec r')
=-{\tilde C(\vec r,\vec r') \over N_{\rm eff} }\,,
\label{arbcovar}
\end{equation}
with
\begin{equation}
N_{\rm eff} = \left[ \sum_{\mu\mu'} |A_{\mu\mu'}|^2 \right]^{-1} \;.
\label{neffgen}
\end{equation}
The definition (\ref{neffgen}) of the effective dimension generalizes the
finite RMT result of Eqs.~(\ref{c2rmt}) and (\ref{covarrmt}) to an arbitrary
shape.  For an ellipse with an aspect ratio $a/b=2$, $N_{\rm eff}$ increases by
a factor of $1.12$ as compared with a circle of the same area; for an ellipse
with an aspect ratio $a/b=4$, $N_{\rm eff}$ increases by a factor of $1.39$.

\begin{figure}[ht] \begin{center} \leavevmode \parbox{0.5\textwidth}
{
\psfig{file=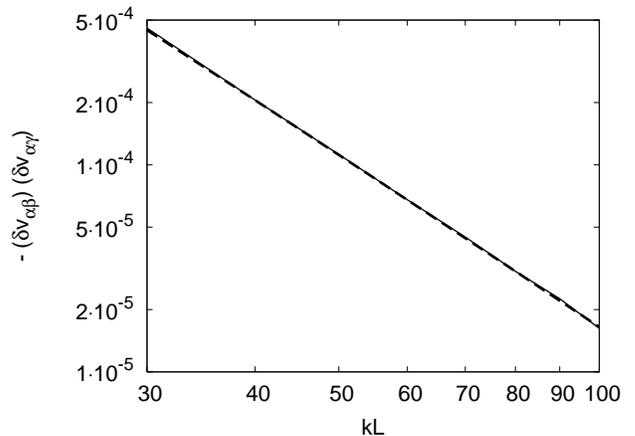,width=0.33\textwidth,angle=270}
}
\end{center}
\protect\caption{
The diagonal interaction matrix element covariance $\overline{\delta
v_{\alpha\beta} \delta v_{\alpha\gamma}}$ (for wave functions $\psi_\beta$ and
$\psi_\gamma$ at the same energy $\hbar^2 k^2/2m$) is plotted versus $kL$ for
real random waves inside a disk.  The result obtained by substituting
Eq.~(\ref{correldiff}) into Eq.~(\ref{covarexpr}) and integrating numerically
(solid line) is compared with the the analytic RMT approximation of
Eq.~(\ref{rmtapprox}) (dashed line).
}
\label{figcovar}
\end{figure}

 In Fig.~\ref{figcovar} we compare the covariance $\overline{\delta
v_{\alpha\beta} \delta v_{\alpha\gamma}}$ calculated numerically in the
normalized random wave model with the RMT expression (\ref{rmtapprox}).
Here we have again used a disk geometry, but very similar results are obtained
in other geometries, e.g., a stadium billiard~\cite{stadium}.  We note the
surprisingly good agreement between the full numerical result and the schematic
RMT approximation.

 The arguments leading to Eq.~(\ref{arbcovar}) generalize to the
case where $\psi_\beta$ and $\psi_\gamma$ are orthogonal random waves at
different energies, as long as the energy difference is classically small,
$\delta k \equiv|k_\beta-k_\gamma| \ll \overline{k} \equiv
(k_\beta+k_\gamma)/2$.  Eq.~(\ref{arbcovar}) for the correlator is
unchanged, provided we use the generalized definition
\begin{equation}
A_{\mu\mu'} = {1 \over V} \int_V d\vec r \, J_\mu(k_\beta r) e^{-i\mu\theta}
J_{\mu'}(k_\gamma r) e^{i \mu' \theta}\,,
\label{agen}
\end{equation}
and $\tilde C(\vec r,\vec r')$ is evaluated at $k=\overline{k}$.  Note that
$A_{\mu \mu'}$, and thus $N_{\rm eff}$ in Eq.~(\ref{neffgen}), are now functions of two wave numbers $k_\beta$ and $k_\gamma$, or equivalently of $\overline{k}$ and $\delta k$. The final result for the covariance becomes
\begin{eqnarray}
\overline{\delta v_{\alpha\beta} \delta v_{\alpha\gamma}} &\approx & -\Delta^2
{3 \over \pi} {\overline{k}L \over N_{\rm eff}(\overline{k},\delta k)} \left({2
\over \beta} \right)^2 {\ln \overline{k}L +b_g\over (\overline{k}L)^3}
\nonumber \\ &+& O\left({\Delta^2 \over (\overline{k}L)^4}\right) \,,
\label{covardk}
\end{eqnarray}
where $N_{\rm eff}$ depends implicitly on both wave vectors $k_\beta$ and
$k_\gamma$ (or alternatively $\bar k$ and $\delta k$) through
Eqs.~(\ref{neffgen}) and (\ref{agen}).

 The energy difference can be written as $E_\beta - E_\gamma
 =2\,\delta k \,\overline{k}\,(\hbar^2/2m)
 \sim (\delta k \,L) E_T$, where $E_T \sim \hbar/(L/v)\sim
\overline{k}\,(\hbar^2/mL)$ is the ballistic Thouless energy.  When this energy
difference is small compared with the Thouless energy, i.e., when the wave
vector difference $\delta k$ is quantum-mechanically small ($\delta k\, L \ll
1$), the ratio $N_{\rm eff}/\overline{k}L$ reduces to a shape-dependent
constant independent of $\overline{k}$.  For a disk, this constant is
$1.85/\sqrt{\pi}$ [see Eq.~(\ref{neff})] and Eq.~(\ref{covardk}) reduces to
Eq.~(\ref{rmtapprox}).

\begin{figure}[ht] \begin{center} \leavevmode \parbox{0.5\textwidth}
{
\psfig{file=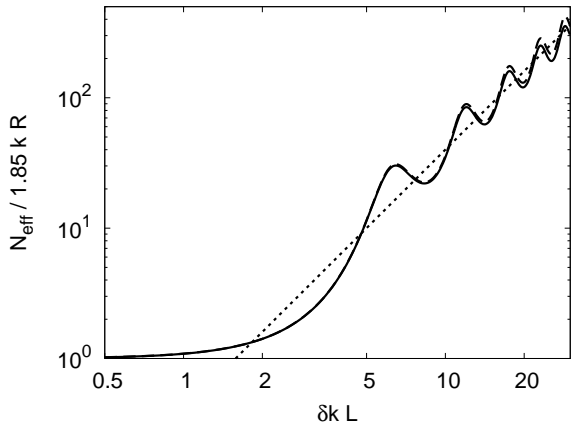,width=0.33\textwidth,angle=270}
}
\end{center}
\protect\caption{
The factor $N_{\rm eff}$, which governs the reduction of the covariance
$\overline{\delta v_{\alpha\beta} \delta v_{\alpha\gamma}}$ as compared with
the variance $\overline{\delta v_{\alpha\beta}^2}$, is computed for a disk
geometry in accordance with the definitions (\ref{neffgen}) and (\ref{agen}).
$N_{\rm eff}$ is computed as a function of $\overline{k}L$ and $\delta k \,L$,
where $\overline{k}=(k_\beta+k_\gamma)/2$, and $\delta k= |k_\beta -
k_\gamma|$.  The solid curve corresponds to $\overline{k}L=30$ and the dashed
curve to $\overline{k}L=70$.  The dotted line shows the quadratic scaling with $\delta
k\,L$, as predicted by Eq.~(\ref{neffscaledkl}).
}
\label{figneffdk}
\end{figure}

 On the other hand, when the energy difference is quantum mechanically large
(i.e., large compared with the Thouless energy) but still classically small ($1
\ll \delta k\,L \ll \overline{k}L$), then the oscillatory functions
$J_\mu(k_\beta r)$ and $J_{\mu'}(k_\gamma r)$ go in and out of phase with each
other $O(\delta k\,L)$ times between $r=0$ and $r \sim L$.  The overlaps
$A_{\mu\mu'}$ as defined by Eq.~(\ref{agen}) are therefore reduced by a factor
$1/\delta k\,L$ (as compared with the case $\delta k\,L \ll 1$), and $N_{\rm eff}$
in Eq.~(\ref{neffgen}) increases
\begin{equation}
N_{\rm eff}(\delta k\,L) \sim (\delta k\,L)^2 \cdot N_{\rm eff}(\delta k\,L=0)
\sim (\delta k\,L)^2 \cdot \overline{k}L \;.
\label{neffscaledkl}
\end{equation}
$N_{\rm eff}/\overline{k}L$ in Eq.~(\ref{covardk}) now becomes $O(\delta k\,L)^{2}$ instead of order unity.
The growth of $N_{\rm eff}$ with increasing $\delta k\,L$ is shown in
Fig.~\ref{figneffdk} in the special case of a disk geometry, for two values of
$\overline{k}L$.  Similar results are obtained for other geometries.

Eq.~(\ref{covardk}) then becomes
\begin{equation}
\overline{\delta v_{\alpha\beta} \delta v_{\alpha\gamma}} \sim
-\Delta^2 \left ({2 \over \beta}\right)^2
 {\ln \overline{k}L +b_g \over (\overline{k}L)^3(\delta k \, L)^2} \,.
\label{covarscale}
\end{equation}

\begin{figure}[ht] \begin{center} \leavevmode \parbox{0.5\textwidth}
{
\psfig{file=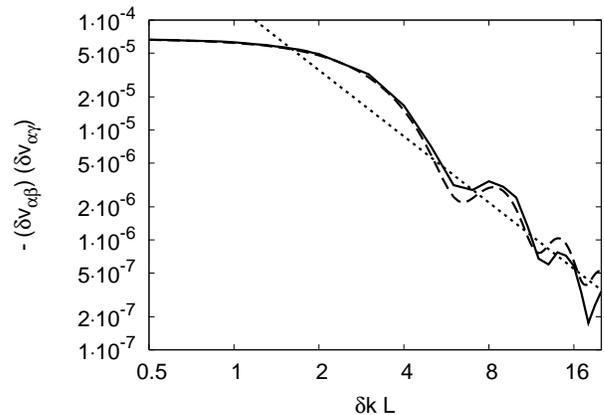,width=0.33\textwidth,angle=270}
}
\end{center}
\protect\caption{
The diagonal interaction matrix element covariance $\overline{\delta
v_{\alpha\beta} \delta v_{\alpha\gamma}}$ for $\delta k \equiv k_\beta -
k_\gamma \ne 0$ is plotted versus $\delta k L \sim \delta E/E_T$ for real
random waves inside a disk; $\overline{k}\, L =(k_\beta+k_\gamma)L/2$ is fixed
at $60$.  The numerical result obtained by substituting Eq.~(\ref{correldiff})
(with $A_{\mu \mu'}$ now given by Eq.~(\ref{agen})) into Eq.~(\ref{covarexpr})
(solid line) is compared with the the analytic RMT approximation of
Eq.~(\ref{covardk}) (dashed line).  The dotted line describes the scaling
$\overline{\delta v_{\alpha\beta} \delta v_{\alpha\gamma}} \sim (\delta
k\;L)^{-2}$ of Eq.~(\ref{covarscale}), valid for $\delta k \; L \gg 1$.
}
\label{figcovardk}
\end{figure}

 In Fig.~\ref{figcovardk} we show the dependence of the diagonal interaction
matrix element covariance on $\delta k L$ for a disk at fixed
$\overline{k}L=60$.  The numerical covariance of the random wave model (solid
line) is in reasonable agreement with the RMT expression (\ref{covardk})
(dashed line).

\section{One-body matrix elements}\label{one-body}

 When an electron is added to the finite dot, charge accumulates on the surface
and its effect can be described by a one-body potential energy ${\cal V}(\vec
r)$.  For an elliptical 2D geometry, the variation in this mean field potential
energy due to the addition of an electron is given by
\begin{equation}
{\cal V}(x,y)= -{\Delta /4 \over \sqrt{1-x^2/a^2-y^2/b^2}} \,,
\label{v1pot}
\end{equation}
where we have used the Thomas-Fermi screening length to obtain the scale of
${\cal V}$.

 We note that for non-elliptical shapes, the correct form of ${\cal V}(\vec r)$
may be calculated numerically.  However, we will see below that the
fluctuations of its matrix elements depend only weakly on the precise form of
the potential, and are instead dominated by the overall normalization and the
$1/\sqrt{d}$ divergence near the boundary.

 The diagonal matrix elements of ${\cal V}(\vec r)$ are given by
\begin{equation}
\label{valphadef}
v_\alpha \equiv {\cal V}_{\alpha \alpha} =\int_V d
\vec r \; |\psi_\alpha(\vec r)|^2 \, {\cal V}(\vec r) \,.
\end{equation}
Using $\overline{|\psi_\alpha(\vec r)|^2} =1/V$, we find
$\overline{v_\alpha}= {1 \over V} \int_V d \vec r \; {\cal V}(\vec r) \equiv
\bar {\cal V} = -\Delta/ 2$.

 Again, of main interest are the fluctuations in $v_\alpha$.  As in
Section~\ref{randvab}, we begin by expressing the variance in terms of the wave
function intensity correlator~\cite{alhassid02,usaj02}
\begin{eqnarray}
\label{v1expr}
\overline{\delta v_\alpha^2} &=& \int_V \int_V d\vec r \, d\vec r' \; {\cal
V}(\vec r) \tilde C(\vec r,\vec r') {\cal V}(\vec r') \\ \label{v1expr1}
&=& \int_V \int_V d\vec r \, d\vec r' \; \tilde {\cal V}(\vec r)
C(\vec r,\vec r') \tilde {\cal V}(\vec r') \,,
\end{eqnarray}
where $\tilde{\cal V}={\cal V}-\overline{\cal V}$. Because
only one power of $C$ or $\tilde C$ appears in the above expressions, the
integral is dominated by distant pairs of points $|\vec r-\vec r'| \sim L$, and
scales as $1/kL$:
\begin{equation}
\overline{\delta v_\alpha^2} ={ c_g \over \beta} {\Delta^2 \over kL}
+O\left({\Delta^2 \over (kL)^2}\right) \,,
\label{v1power}
\end{equation}
where $c_g$ is a shape-dependent dimensionless coefficient.

\begin{figure}[ht] \begin{center} \leavevmode \parbox{0.5\textwidth}
{
\psfig{file=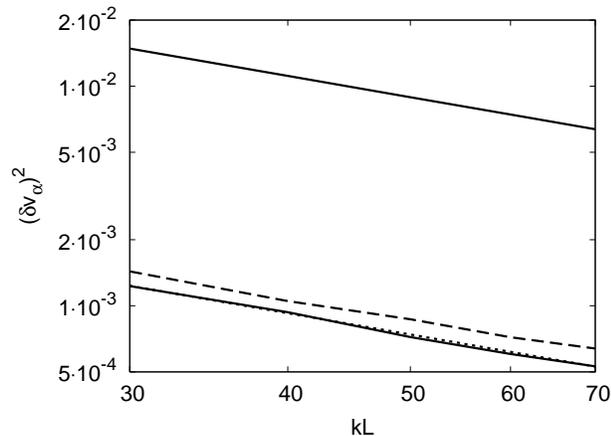,width=0.33\textwidth,angle=270}
}
\end{center}
\protect\caption{
The expression (\ref{v1expr}) for the one-body matrix element variance
$\overline{\delta v_\alpha^2}$ is plotted as a function of $kL$ for real random
waves inside a disk using the analytic potential of Eq.~(\ref{v1pot}) (lower
solid curve) and the schematic potential of Eq.~(\ref{v1fakepot}) (lower dashed
curve).  The dotted line, which is almost indistinguishable from the lower
solid curve, is the power-law prediction of Eq.~(\ref{v1power}).  The upper
solid curve is calculated from an unsubtracted integral, where the unnormalized
propagator $C(\vec r,\vec r')$ is substituted for $\tilde C(\vec r,\vec r')$ in
Eq.~(\ref{v1expr}).
}
\label{figv1rand}
\end{figure}

 In Fig.~\ref{figv1rand} we show the result of integrating Eq.~(\ref{v1expr})
for a disk with $\beta=1$.  The solid lower curve corresponds to the analytic
potential of Eq.~(\ref{v1pot}) while the dashed lower curve corresponds to the
``schematic" potential
\begin{equation}
{\cal V}_{\rm sch}(\vec r) \sim \left(\min_{\vec R \in {\rm {\cal C}}}|\vec
r-\vec R|\right)^{-{1 \over 2}}\,, \label{v1fakepot}
\end{equation}
which has the same singularity at the boundary ${\cal C}$ as the true potential
(\ref{v1pot}) and is scaled to have the same average as ${\cal V}(\vec r)$.
The upper solid curve is obtained by substituting the unnormalized correlator
$C(\vec r,\vec r')$ in place of $\tilde C(\vec r,\vec r')$ in
Eq.~(\ref{v1expr}).  We note that normalization, which had only a moderate
effect on the two-body matrix element fluctuations, here reduces the variance
by a full order of magnitude, resulting in a very small prefactor $c_g$ in
Eq.~(\ref{v1power}).  (For a disk, $c_g=0.035$, and the power-law prediction of
Eq.~(\ref{v1power}) is indicated in Fig.~\ref{figv1rand} as a dotted line.)
This reduction in the variance is due to the fact that after normalization,
according to Eq.~(\ref{corrnorm}), the one-body integrand (\ref{v1expr}) ceases
to be positive everywhere, resulting in substantial cancellation in the
integral.  This is in contrast with the two-body integrand in
Eq.~(\ref{vabint2}), which remains positive even after normalization.  On the
other hand, the difference between the real and the schematic potential after
normalization is only an $18\%$ effect.  This fact is useful for studying one-body matrix elements in chaotic geometries, where no analytic form exists for the surface-charge potential.

 The small value of the coefficient $c_g$ leads to
values of $\overline{\delta v_\alpha^2}$  (see Fig.~\ref{figv1rand}) that are numerically smaller in the physically interesting $kL$ regime than the corresponding values of the two-body matrix element variance $\overline{\delta v_{\alpha\beta}^2}$ (see Fig.~\ref{figrandvab}), despite the fact that the former is parametrically
larger in a $1/kL$ expansion.

 The two lower curves in Fig.~\ref{figrandvabshape} show the dependence of $\overline{\delta v_\alpha^2}$ on the dot's geometry.
We find the one-body matrix element variance to
have much stronger shape dependence than the two-body matrix element variance.
This is due mostly to the fact that here the coefficient of the leading term in
Eq.~(\ref{v1power}) is already shape dependent, in contrast with the
shape-independence of the leading logarithmic term in the two-body expression
(\ref{leadingvabrand}).  In particular, an aspect ratio of $r=4$ results in a
$14\%$ reduction in the coefficient $c_g$ in Eq.~(\ref{v1power}).  We also note
the divergence between the $kL=30$ and $kL=70$ curves for $r>4$, indicating
that for very large aspect ratios, the simple inverse relationship between $kL$
and the variance breaks down in the energy range of interest and the subleading
terms in Eq.~(\ref{v1power}) acquire greater importance.

\section{Matrix element distributions}
\label{secdistr}

 The central limit theorem implies that all interaction matrix elements such as
$v_{\alpha\beta}$, $v_{\alpha\alpha}$, $v_{\alpha\beta\gamma\delta}$, and
$v_\alpha$ in the random wave model must be distributed as Gaussian random
variables as $kL \to \infty$, since each of them is defined as an integral over
a large volume of a random integrand with decaying correlations.  This
justifies our focus so far on the variance and covariance of these matrix
elements.  However, we have seen above that
non-universal finite $kL$ effects can be significant in the
experimentally relevant regime $kL \le 70$, most notably for the variance
ratios shown in Fig.~\ref{figrandratios}.  Here we look for finite-$kL$ effects
on the shape of matrix element distributions.

\begin{figure}[ht] \begin{center} \leavevmode \parbox{0.5\textwidth}
{
\psfig{file=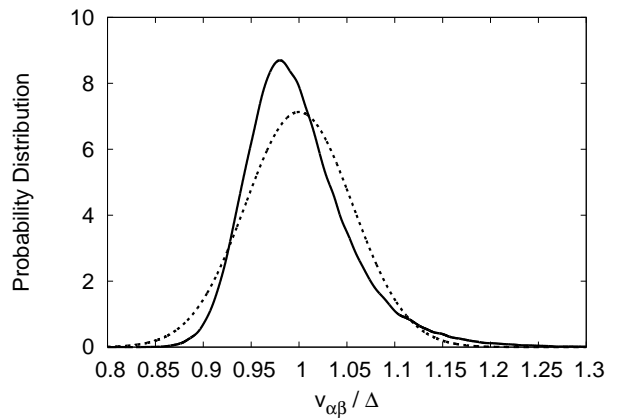,width=0.33\textwidth,angle=270}
}
\end{center}
\protect\caption{
The distribution of diagonal interaction matrix elements $v_{\alpha\beta}$ is
shown for normalized real random waves in a disk at $kL=70$ (solid curve).
A Gaussian distribution with the same mean
and variance is shown as a dotted curve for comparison.
}
\label{figdistr}
\end{figure}

 The distribution of diagonal interaction matrix element $v_{\alpha\beta}$ is
shown in Fig.~\ref{figdistr} for normalized random waves in a disk (solid curve).  Results for other dot's geometries and for other matrix elements (i.e., $v_{\alpha\alpha}$, $v_{\alpha\beta\gamma\delta}$, and $v_{\alpha}$) are qualitatively similar.  We observe that the distribution has a long tail on the right side as compared with a Gaussian distribution of the same mean and variance (dotted curve).  In other words, there is an excess of anomalously large matrix elements, compensated for by a reduction in the median to a value slightly below $\Delta$.  Indeed, the right tail has a shape closer to an exponential than to a Gaussian, reminiscent of the distribution of many-body matrix elements in realistic atomic shell model calculations~\cite{flambaum}.

 Deviations from a Gaussian shape can be quantified by considering higher
moments such as the skewness
\begin{equation}
\gamma_1 = { \overline{\delta v_{\alpha\beta}^3} \over
\;\;\;\;\left[\;\overline{\delta v_{\alpha\beta}^2}\;\right]^{3/2} }
\label{skewdef}
\end{equation}
and excess kurtosis
\begin{equation}
\gamma_2 = { \overline{\delta v_{\alpha\beta}^4} - 3
 \left[\; \overline{\delta v_{\alpha\beta}^2}\;\right]^2
\over
\left [\overline{\delta v_{\alpha\beta}^2}\; \right] ^2 } \,.
\label{kurtdef}
\end{equation}
For large $kL$, we may estimate these higher moments in a manner analogous
to our estimate for the variance in Eq.~(\ref{vabint}).  Specifically,
\begin{eqnarray}
\label{skew}
&\overline{\delta v_{\alpha\beta}^3} \equiv & \nonumber \\[6pt] \equiv &
\Delta^3 V^3 \overline{\left[\int_V d\vec r \; |\psi_\alpha(\vec r)|^2
|\psi_\beta(\vec r)|^2
- \overline{\int_V d\vec r \; |\psi_\alpha(\vec r)|^2
|\psi_\beta(\vec r)|^2} \right ]^3} &
\nonumber \\
& = \Delta^3 V^3  \Bigl\{
\int_V \int_V \int_V d\vec r \,d\vec r' \, d\vec r'' \nonumber & \\[6pt] &
\overline{\left [ (|\psi_\alpha(\vec r)|^2 -{1 \over V})
(|\psi_\alpha(\vec r')|^2 -{1\over V})
(|\psi_\alpha(\vec r'')|^2 -{1\over V}) \right ]} \nonumber & \\[6pt] &
\overline{\left [ (|\psi_\beta(\vec r)|^2 -{1 \over V})
(|\psi_\beta(\vec r')|^2 -{1\over V})
(|\psi_\beta(\vec r'')|^2 -{1\over V}) \right ]}
& \nonumber \\ & + \; \cdots\; \Bigr\} & \nonumber \\
& \approx \Delta^3 V^3
\int_V \int_V \int_V d\vec r \,d\vec r' \, d\vec r''
\overline{u_\alpha(\vec r)u_\alpha(\vec r') u_\alpha(\vec r'')}^2 \,
\end{eqnarray}
where $u_\alpha(\vec r)=|\psi_\alpha(\vec r)|^2-{1 \over V}$ is the excess
intensity.  The ellipsis in Eq.~(\ref{skew})
indicates omitted terms involving the intensity correlation of two different
wave functions at different points, such as the correlator $C_2(\vec r,\vec
r')$ defined in Eq.~(\ref{twoeig}) and its generalization to three distinct
points.

 The three-point intensity correlator for a single random wave function can be
computed to leading order in $kL$ by following the same procedure that led to
the leading-order two-point intensity correlator (\ref{corr}): performing all
possible contractions to rewrite the quantity of interest as a product of
amplitude correlators and making use of Eq.~(\ref{ampcorr}).  Thus
\begin{eqnarray}
& &\overline{u_\alpha(\vec r)u_\alpha(\vec r') u_\alpha(\vec r'')} =
\nonumber \\[3pt] &=&
\overline{\left(|\psi(\vec r)|^2-{1 \over V}\right)
\left(|\psi(\vec r')|^2-{1 \over V}\right)
\left(|\psi(\vec r'')|^2-{1 \over V}\right)} \nonumber \\
& \approx & c_{3\beta} \;\,
\overline{\psi^\ast(\vec r)\psi(\vec r')} \;\;
\overline{\psi^\ast(\vec r')\psi(\vec r'')} \;\;
\overline{\psi^\ast(\vec r'')\psi(\vec r)} \nonumber \\[3pt]
&=&{c_{3\beta} \over V^3}
J_0(k|\vec r-\vec r'|)
J_0(k|\vec r'-\vec r''|)
J_0(k|\vec r''-\vec r|) \;,
\label{3ptcorr}
\end{eqnarray}
where $c_{31}=8$ for real waves and $c_{32}=2$ for complex waves are
combinatorial factors.  Normalization effects analogous to those
included in Eq.~(\ref{corrnorm}) are subleading in this case, since
$\overline{u_\alpha(\vec r)u_\alpha(\vec r') u_\alpha(\vec r'')}$ is a
fluctuating quantity with average close to zero instead of being everywhere
positive.  The result (\ref{3ptcorr}) scales as $V^{-3}(kL)^{-3/2}$ in the
typical case when the inter-point separations are all of order $L$.

 Combining Eq.~(\ref{3ptcorr}) with Eq.~(\ref{skew}), we have
\begin{eqnarray}
\overline{\delta v_{\alpha\beta}^3}&=&
c_{3\beta}^2\left({\Delta\over V}\right)^3
\int_V \int_V \int_V d\vec r \,d\vec r' d\vec r'' \{
J_0^2(k|\vec r-\vec r'|)  \nonumber \\
& & \times J_0^2(k|\vec r'-\vec r''|)
J_0^2(k|\vec r''-\vec r|) \} \label{skewint} \\
&=& b_{3g} c_{3\beta}^2 {\Delta^3 \over (kL)^3} \,,
\label{skewscaling}
\end{eqnarray}
where $b_{3g}$ is a shape-dependent constant analogous to $b_g$ in the
calculation of the variance (\ref{bcoeff}).  In contrast to the variance calculation (\ref{leadingvabrand}), here we have no large short-distance contribution resulting in a logarithmic contribution at leading order.  The irrelevance of the short-distance contribution may be seen by noting that a fraction $\sim \epsilon^4$ of the integration space in Eq.~(\ref{skewint}) has points $\vec r$, $\vec r'$, and $\vec r''$ all within distance $\epsilon$ of one another, while the integrand is only enhanced by a factor $\sim \epsilon^{-3}$ in this region.  For a disk geometry, $b_{3g}=1.3$.

 Similarly, the fourth moment $\overline{\delta v_{\alpha\beta}^4}$ involves the four-point intensity correlation function, which to
leading order in $1/kL$ takes the form
\begin{eqnarray}
& \overline{u_\alpha(\vec r)u_\alpha(\vec r') u_\alpha(\vec r'')
u_\alpha(\vec r''')} \approx & \nonumber \\[3pt]
& c_{4\beta}\big [
J_0(k|\vec r\!-\!\vec r'|)
J_0(k|\vec r'\!-\!\vec r''|)
J_0(k|\vec r''\!-\!\vec r'''|)
J_0(k|\vec r'''\!-\!\vec r|) & \nonumber \\[3pt] & +
(\vec r' \leftrightarrow \vec r'') +
(\vec r'' \leftrightarrow \vec r''')\big] & \nonumber \\
& +\left ({2 \over \beta}\right )^2
\big[J_0^2(k|\vec r-\vec r''|)J_0^2(k|\vec r'-\vec r'''|) & \nonumber
\\ & +(\vec r' \leftrightarrow \vec r'') +
(\vec r'' \leftrightarrow \vec r''')\big] \,. &
\label{u4}
\end{eqnarray}
Here $c_{41}=16$  and $c_{42}=2$ are combinatorial factors.  Performing a four-fold integral over volume, we obtain
\begin{eqnarray}
\overline {\delta v_{\alpha\beta}^4}
&\approx &
\Delta^4 V^4
\int_V \!\int_V \!\int_V \!\int_V \! d\vec r \,d\vec r' \, d\vec r'' \, d\vec r''' \;
\nonumber \\
& & \times \overline{u_\alpha(\vec r)u_\alpha(\vec r') u_\alpha(\vec r'')
u_\alpha(\vec r''')}^2 \nonumber \\[3pt]
& \approx &
3 \left(c_{4\beta}^2 + \left({2 \over \beta}\right)^4 \right)
\left ({\Delta \over V}\right )^4 \int_V \!\int_V \!\int_V \!\int_V \! d\vec r \,d\vec
r' \, d\vec r'' \, d\vec r''' \; \nonumber \\
& &  \big[J_0^2(k|\vec r-\vec r'|)
J_0^2(k|\vec r'-\vec r''|) \nonumber \\[3pt]
& & \times J_0^2(k|\vec r''-\vec r'''|)
J_0^2(k|\vec r'''-\vec r|)\big]
+ 3\; \overline {\delta v_{\alpha\beta}^2}^2 \;,
\label{dv4}
\end{eqnarray}
where in the second step we have omitted terms in the integrand containing odd
powers of $J_0$ (these terms have oscillating sign and contribute to the
integral only at subleading order in $kL$).  We have also separated out a term
proportional to the square of the variance (\ref{vabint}); this term
corresponds to disconnected diagrams and does not contribute to the fourth
cumulant or the excess kurtosis $\gamma_2$.  The final result is
\begin{equation}
\overline {\delta v_{\alpha\beta}^4}
- 3\; \overline {\delta v_{\alpha\beta}^2}^2 =
3\, b_{4g}\left(c_{4\beta}^2 + \left({2 \over \beta}\right)^4 \right)
{\Delta^4 \over (kL)^4}\,,
\label{kurtscaling}
\end{equation}
where the shape-dependent coefficient $b_{4g}=1.0$ for a disk geometry.
Higher-order cumulants beyond (\ref{skewscaling}) and (\ref{kurtscaling}) may
be computed similarly.  It is evident from the preceding discussion that the
$n-$th cumulant involves an $n-$fold integral of a product of $2n$ Bessel
functions and scales as $\Delta^n/(kL)^n$, with a combinatorial
$\beta-$dependent prefactor and a geometry-dependent overall dimensionless
constant.

 When the cumulants $\overline{\delta v_{\alpha\beta}^3}$ and $\overline {\delta
v_{\alpha\beta}^4}
- 3 \,\overline {\delta v_{\alpha\beta}^2}^2$ for normalized random waves in a disk are
  computed numerically, they compare well with the power-law predictions of
(\ref{skewscaling}) and (\ref{kurtscaling}) at sufficiently large $kL$ (not shown).

\begin{figure}[ht] \begin{center} \leavevmode \parbox{0.5\textwidth}
{
\psfig{file=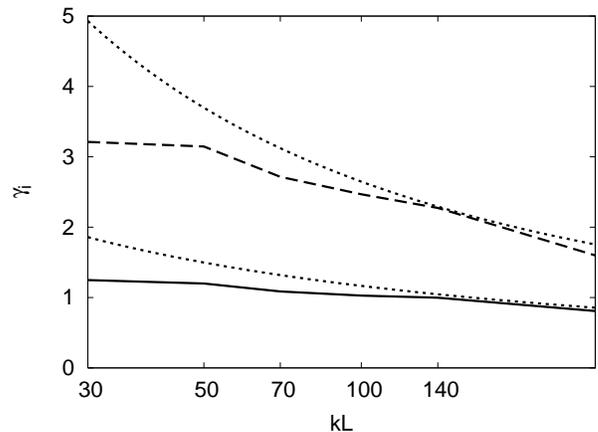,width=0.33\textwidth,angle=270}
}
\end{center}
\protect\caption{
The skewness $\gamma_1$ (\ref{skewdef}), indicated by a solid curve, and excess
kurtosis $\gamma_2$ (\ref{kurtdef}), indicated by a dashed curve, for the
distribution of diagonal interaction matrix elements $v_{\alpha\beta}$ in a
disk are computed within the normalized real random wave model.  Analytic predictions
(\ref{gam3}) and (\ref{gam4}) valid at large $kL$ are indicated by dotted
lines.
}
\label{fighighmom2}
\end{figure}

 Combining Eqs.~(\ref{skewscaling}) and (\ref{kurtscaling}) with our previous
result (\ref{leadingvabrand}) for the variance, we find the skewness
\begin{equation}
\gamma_1 = b_{3g} \; c_{3\beta}^2 \left({\beta \over 2} \right)^3
\left({\pi \over 3}\right )^{3/2} (\ln kL)^{-3/2}
\label{gam3}
\end{equation}
and excess kurtosis
\begin{equation}
\gamma_2 = b_{4g}\left(c_{4\beta}^2 + \left({2 \over \beta}\right)^4 \right)
\left({\pi^2 \over 3}\right) (\ln kL)^{-2}
\label{gam4}
\end{equation}
at large $kL$.  Because the decay is only logarithmic, $\gamma_1$ and
$\gamma_2$ never become small for values of $kL$ relevant in the experiments,
as seen in Fig.~\ref{fighighmom2}.  The same holds for the
higher moments.  Therefore, the random wave model leads to strongly non-Gaussian matrix element distributions.

 Higher cumulants for the distribution of double-diagonal interaction matrix
elements $v_{\alpha\alpha}$ and for off-diagonal matrix elements
$v_{\alpha\beta\gamma\delta}$ may be obtained similarly; only the
$\beta$-dependent combinatorial prefactors in Eqs.~(\ref{skewscaling}) and
(\ref{kurtscaling}) are modified and the $n$-th cumulants again scale as
$\Delta^n/(kL)^n$.  Of course, the skewness $\gamma_1$ and all odd moments
vanish identically for $v_{\alpha\beta\gamma\delta}$, since the distribution in
this case is manifestly symmetric around zero.

 Finally, we find that the distribution of one-body matrix elements $v_\alpha$
 approaches more rapidly a Gaussian form, with skewness and excess kurtosis
decaying to zero as a power law instead of a logarithm in $kL$.  For
example, to leading order in $1/kL$,
\begin{eqnarray}
\label{v1third}
\overline{\delta v_\alpha^3} &=& {1 \over V^3} \int_V \int_V d\vec r \, d\vec r'
\, d\vec r''\; \bigg[{\cal V}(\vec r) J_0(k|\vec r-\vec r'|) {\cal V}(\vec r')
\nonumber  \\ &\times &J_0(k|\vec r-\vec r''|) {\cal V}(\vec r'') J_0(k|\vec r''-\vec
r|)\bigg] \sim  {\Delta^3 \over (kL)^2 }
\end{eqnarray}
and comparing with Eq.~(\ref{v1power}) we find
\begin{equation}
\gamma'_1  \sim (kL)^{-1/2} \,,
\end{equation}
where $\gamma'_1$ is defined as in (\ref{skewdef}) with $v_\alpha$ replacing
$v_{\alpha\beta}$.  At the experimentally relevant values of $kL$, the deviation
from Gaussian behavior for one-body matrix elements is nevertheless
significant, though it is less pronounced than for two-body matrix elements.
Even at $kL=70$, we have $\gamma'_1 \approx \gamma'_2\approx 0.6$.

\section{Summary and conclusion}
\label{summary}

 We have studied fluctuations of two-body interaction matrix elements and surface charge one-body matrix elements in ballistic quantum dots as a function of semiclassical parameter $kL$. Understanding the quantitative behavior of these fluctuations is important for a proper analysis of peak spacing statistics and scrambling effects in the Coulomb blockade regime.

 The variance and higher cumulants of two-body and one-body
matrix elements can be derived from spatial correlations of
the single-particle Hartree-Fock wave functions.  For a chaotic dot, we have estimated these correlations to leading order in $kL$ using
Berry's random wave model.  The variances of two-body matrix elements are found  to scale as $\ln kL/ (kL)^2$, with universal prefactors that depend only on the symmetry class of the system.  Geometry-dependent effects on the variance enter at the order of $1/(kL)^2$, where the random wave intensity correlator must be corrected
to satisfy individual wave function normalization in a finite volume. To understand such corrections we have studied a normalized random wave model. Variance ratios such as $\overline{\delta v_{\alpha\alpha}^2}/\overline{\delta v_{\alpha\beta}^2}$ converge only with a logarithmic rate in the $kL \to \infty$ limit. As a result, the asymptotic values of such ratios are not yet reached in the regime of experimental interest.  The interaction matrix element covariance is important in
calculations of spectral scrambling; this quantity, which must be negative on
average due to a sum rule, is computed as a function of energy separation
using the intensity correlator between two orthonormal random wave
functions.

The variance of one-body matrix elements $v_{\alpha}$ (of, e.g., the surface charge potential) is affected by normalization even at leading order, resulting in $O(1/kL)$ scaling in a random wave model, with a shape-dependent prefactor that may be computed using the normalized intensity correlator. For typical experimental values of $kL$, we find the distributions of two-body and one-body matrix elements to be strongly non-Gaussian.  Thus, higher cumulants of these matrix elements may play a role in the peak spacing statistics, especially in the case of two-body matrix elements where we have seen that the approach to a Gaussian distribution is logarithmically slow.

The absence of bimodality in the measured peak spacing distribution at low temperatures cannot be explained by the exchange interaction alone and must originate in the non-universal part of the electron-electron interaction. Sufficiently large fluctuations of such interaction matrix elements are required to wash out the bimodality of the peak spacing distribution. The fluctuation width estimates derived here in the framework of the random wave model are too small.  It is therefore necessary to study interaction matrix element fluctuations in real chaotic systems and in system with mixed dynamics, where dynamical effects may lead to enhanced fluctuations~\cite{ullmo03}. Such studies are important for the quantitative understanding of spectral scrambling and the measured peak spacing distribution.

\section*{Acknowledgments} We acknowledge useful discussions with Y.~Gefen,
Ph.~Jacquod, and C.\ H.~Lewenkopf.  This work was supported in part by the U.S. Department of Energy Grants No.\ DE-FG03-00ER41132 and DE-FG-0291-ER-40608 and by the National Science Foundation under Grant No.\ PHY-0545390.


\begin{thebibliography}{99}
\bibitem{geometry}
Since the shape dependence is observed to be very weak within the random wave
model, the values of $b'_g$ and $b''_g$ for other geometries are very similar.
\end{thebibliography}
\end{document}